\documentclass[twocolumn,aps,prl,superscriptaddress,showpacs,papersize=a4paper,amsmath]{revtex4-1}
\usepackage[latin9]{inputenc}
\usepackage{float}
\usepackage{mathtools}
\usepackage{amsmath}
\usepackage{amssymb}
\usepackage{graphicx}

\makeatletter
\usepackage{subcaption}
\usepackage{etoolbox}
\usepackage{amsfonts}

\makeatother

\begin{document}
\title{Theory of superconductivity due to Ngai's mechanism in lightly doped
SrTiO$_{3}$}
\author{D.~E.~Kiselov}
\affiliation{Moscow Institute of Physics and Technology}
\affiliation{L.D.Landau Institute for Theoretical Physics, Chernogolovka, Moscow Region, 143432, Russia}
\author{M.~V.~Feigel'man}
\affiliation{L.D.Landau Institute for Theoretical Physics, Chernogolovka, Moscow
Region, 143432, Russia}
\affiliation{Moscow Institute of Physics and Technology}
\begin{abstract}
We develop a theory of superconducting pairing in low-density Strontium
titanate due to quadratic coupling of electron density
to soft transverse optical phonons~\cite{Ngai1974}. It
leads to static attractive potential between electrons which decay length
$l_{\mathrm{eff}}$ that
scales inversely with soft optical gap $\omega_{T}$. For low electron
densities $n\leq10^{18}cm^{-3}$ attraction between electrons is local 
and transition temperature $T_{c}$ was found using Ref.~\cite{Gorkov60}.
The $T_{c}(n)$ dependence in agreement with experimental data~\cite{Behnia2014}
for low doping was calculated. Next, we show that suppression of $T_{c}$ by hydrostatic
pressure~\cite{Rowley2020} and strong increase of $T_{c}$ due to
isotop substitution $^{16}O\to^{18}O$ observed in~\cite{Dirk2016}
are explained within our theory.

\end{abstract}
\date{\today}

\maketitle
\textit{Introduction}. Strontium titanate (STO) is a wide-gap band
insulator known for more than half a century for its unusual properties
related to its proximity to a ferroelectric transition, see Refs~\citep{review1,review2,review3}
for recent reviews. It can be transformed to a very dilute metal by
tiny doping, either by oxygen deficiency or by substitution of small
portion of Ti atoms with Nb. The key feature of this metal which makes
it very different from majority of various metals and doped semiconductors
is that Coulomb interaction is nearly absent due to extremely high
low-temperature static dielectric permeability $\epsilon_{0}\approx20000$.
It results in the huge value of effective Bohr radius, $a_{B}\approx600$nm,
to be compared with normal values $\sim0.1$ nm for usual metals.
Therefore, Coulomb interaction being the major one in metals normally,
appears to be nearly irrelevant for a very dilute STO-based conductor.
Suprisingly enough, lighly doped STO becomes superconducting in a
wide range of conduction electron densities, $10^{17}{\mathrm{c}m}^{-3}\leq n\leq10^{20}{\mathrm{c}m}^{-3}$.

Although first report on superconductivity in STO is dated as early
as 1964, see Ref.~\citep{SCfirst}, the mechanism of electron pairing
is under active debates till now (examples can be found in Refs.~\citep{review1,review2,review3}).
The key feature of superconductivity in STO (which makes it very different
from most of the known superconductors) is due to its very low Fermi
energy $E_{F}$ which is much less than Debye energy $\hbar\omega_{D}\approx100meV$.
In result, classical theory of superconductivity based upon Migdal-Eliashberg
equations~\cite{EliashbergTheory}, cannot be applied here. An attempt
to surcumvent this problem was made in Ref.~\cite{RuhmanLee2016}
where pairing due to exchange by very soft plasmons was proposed.
However, the results obtained for $T_{c}$ were found strongly off
the data, if measured value of dielectric constant $\epsilon_{0}\approx20000$
to be used. 
Proximity of STO to ferroelectric critical point is surely the crucial
feature of this material. The paper~\cite{Edge2015} pointed out
the relevance of this criticality to superconductivity; specifically,
they predicted strong increase of $T_{c}$ upon isotop substitution
$^{16}O\to^{18}O$, since the latter is known to produce ferroelectricity
when about $\frac{1}{3}$ of O atoms is replaced by heavy Oxygen~\cite{O18}.
Such an effect was indeed observed soon~\cite{Dirk2016}: 35$\%$
substitution $^{16}O\to^{18}O$ increases $T_{c}$ by factor $\sim1.5$
and also increases upper critical field nearly twice. An important
step forward was made in Ref.~\cite{Dirk2019} which put support
to the early idea~\cite{Ngai1974} about relevance of coupling between
electron density and two soft transverse optical (TO) phonons, those
existence was known since Ref.~\cite{YamadaShirane}. Namely, Ref.~~\cite{Dirk2019}
provide arguments based upon analysis of optical absorption spectra
in favor of large magnitude of this electron coupling to two TO phonons,
of the form $\psi^{\dagger}\psi\mathbf{u}^{2}$, where $\psi$ is
the electron annihilation operator and $\mathbf{u}$ is the TO phonon
displacement amplitude. Very low gap known for these phonons at low
temperatures, $\hbar\omega_{T}\sim1.5meV$, is directly related to
large $\epsilon_{0}$ value, $\omega_{T}\propto1/\sqrt{\epsilon_{0}}$.

In the present Letter we further develop the ideas proposed in Refs.~\cite{Ngai1974,Dirk2019}
and use quadratic coupling between electrons and TO phonons as phenomenological
input for our theory. We concentrate on the lowest-density limit $n<n_{c1}\approx1.5\cdot10^{18}cm^{-3}$
where single-band Fermi-liquid in realized~\cite{Behnia2014}, and
demonstrate that electron-electron interaction mediated via two TO
phonons leads to consistent description of superconducting $T_{c}(n)$
evolution with $n$, Ref.~\cite{Behnia2014}, and of its giant isotop
effect~\cite{Dirk2016}. Pairing by two-phonon exchange differs a
lot from usual single-phonon exchange but leads to a simple picture:
electrons attract each other via static potential which decays with
a distance as $-V(r)\propto r^{-3}e^{-2r/l_{0}}$, where $l_{0}=s/\omega_{TO}\approx3.3$
nm is the characteristic length related to soft polarization TO mode;
for the  velocity of the TO mode we use $s=7.5\cdot10^{5}cm/s$, see~\cite{sound_velocity}.
At low electron densities we consider, $k_{F}l_{0}/2<0.6$ and electron-electron
scattering can be considered short-range. Frequency dispersion of
e-e scattering occurs then at the energy scale $\epsilon\geq\omega_{T}$,
and it is relatively weak in the low-density region, since $E_{F}(n)$
does not exceed $\omega_{T}$ considerably.

Superconductivity in a bulk Fermi-gas with local attraction was studied
theoretically long ago by Gor'kov and Melik-Barhudarov~\cite{Gorkov60},
see also more recent paper~\cite{Gorkov2016}. They found expression
for $T_{c}$ similar to the one known for usual BCS theory, with the
major exception that Debye energy is replaced by Fermi energy in the
prefactor, $T_{c}\approx0.27E_{F}e^{-1/\lambda_{0}}$. Dimensionless
coupling constant $\lambda_{0}=\nu_{0}V_{0}$, where $V_{0}$ is the
renormalized electron-electron scattering potential in the $l=0$
scattering state (s-wave), and $\nu_{0}$ is the density of states
at Fermi-level per single spin projection. Below we demonstrate that
attractive short-range potential does indeed appear due to two TO
phonon exchange, and calculate $T_{c}$ as function electron density
$n$. Then we extend our results to the case of partial isotopic substitution
$^{16}O\to^{18}O$, which reduces $\omega_{T}$ and thus increases
$V_{0}$ and $T_{c}$.

\emph{Electron-electron interaction mediated by a pair of TO phonons.}
We start with the action for coupled electron-phonon system close
to ferroelectric transition: 
\begin{eqnarray}
S & = & S_{e}^{(0)}+S_{ph}^{(0)}+S_{\mathrm{int}}\label{action1}\\
S_{ph}^{(0)} & = & \frac{\rho_{m}}{2}\int d^{3}xdt\left[\dot{u}_{\alpha}^{2}-s^{2}(\nabla_{\beta}u_{\alpha})^{2}-\omega_{T}^{2}u_{\alpha}^{2}\right]\nonumber \\
S_{\mathrm{int}} & = & -g\rho_{m}\int d^{3}xdt(\bar{\psi}\psi)u_{\alpha}^{2}\nonumber 
\end{eqnarray}
and $S_{e}^{(0)}$ is just the action of free electron gas with effective
mass $m_{e}=1.8m_{0}$, according to the data from Ref.~\cite{Behnia2013}
for low electron densities in STO, $m_{0}$ being free electron mass.
Here $u_{\alpha}$ is displacement coordinate for TO soft optical
phonon, $\psi(x)$ is the electron field operator and $\rho_{m}=5.11g/cm^{3}$
is the mass density of STO. The action $S_{ph}^{(0)}$ describes long-wave-length
TO phonons with momenta $q\ll K_{BZ}$, where $K_{BZ}$ is the boundary
of the Brilliuen zone. Whenever high-$q$ cut-off will be needed in
the further calculations, we introduce it by using the simplest TO
phonon lattice spectrum of the form appropriate for cubic lattice,
with the BZ including $p_{x,y,z}\in(-\frac{\pi}{a},\frac{\pi}{a})$:
\begin{equation}
\omega^{2}(p)= \omega_T^2+ \frac{4s^{2}}{a^{2}}\left(\sin^{2}\frac{p_{x}a}{2}+\sin^{2}\frac{p_{y}a}{2}+\sin^{2}\frac{p_{z}a}{2}\right)\label{TOspectrum}
\end{equation}
where $a\approx0.4nm$ is the lattice spacing. The coupling constant
$g$ in Eq.(\ref{action1}) has natural dimension ${\tt {Length^{3}/Time^{2}}}$;
we represent it in the form 
\begin{equation}
g=\lambda a^{3}\omega_{L}^{2}\label{g}
\end{equation}
where $\hbar\omega_{L}\approx0.1eV$ is the largest longitidinal optical
gap of STO, and $\lambda\sim1$ is the dimensionless coupling constant
of the problem. In a recent paper~\cite{Nazaryan} the same type
of electron coupling to TO phonons was employed to study theoretically
high-temperature transport properties of lightly doped STO; comparison
of this theory predictions with the data ~\cite{Behnia2020} provide
the value of the coupling constant $\lambda\approx 0.9$. 

\begin{figure}[H]
\includegraphics[width=1\linewidth,height=0.2\textheight]{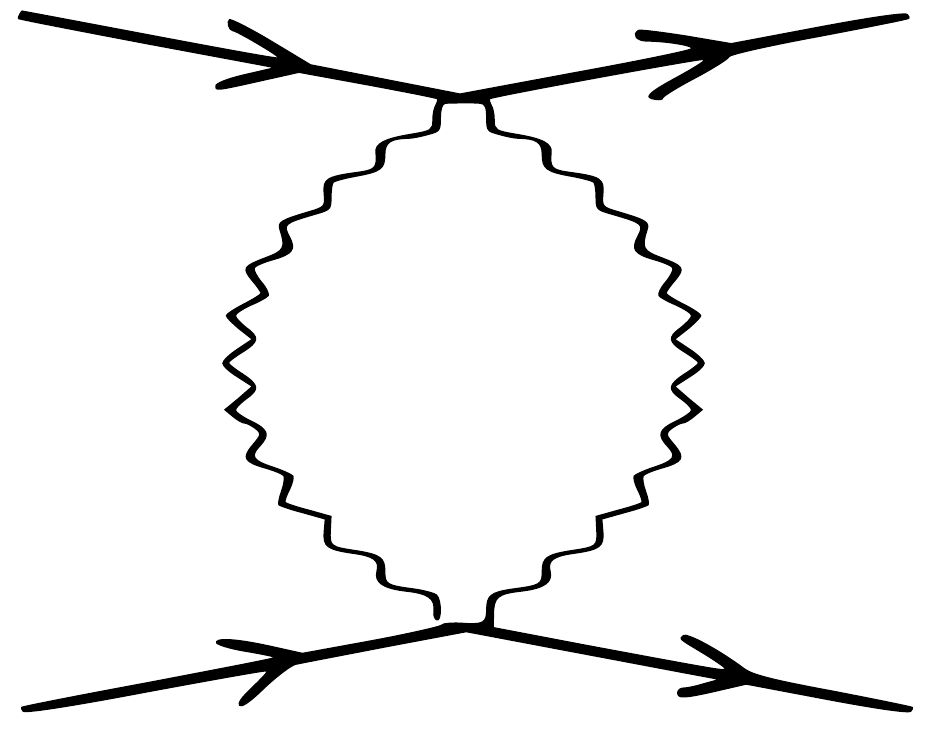}
\caption{\label{fig:diagram} The diagram corresponding to  the static interacting potential between electrons.
Solid (wavy) lines denote electron (phonon) Green functions correspondingly.}
\end{figure}

Static interacting potential between two electrons can be obtained
from the action (\ref{action1}) by integrating out the phonons; the
corresponding diagram is shown in Fig.\ref{fig:diagram}. 

Its analytical expression in Fourrier space is 
\begin{equation}
V(\omega,q)=-4\,g^{2}\rho_{m}^{2}\int\frac{d\epsilon}{2\pi}\int_{BZ}\frac{d^{3}p}{(2\pi)^{3}}D_{0}(\epsilon,\mathbf{p}+\mathbf{q})D_{0}(\omega-\epsilon,\mathbf{p})
\label{Vq}
\end{equation}
where momentum integration goes over simple cubic Brilliuen zone and
phonon Green function $D_{0}$ is given, in the imaginary-time formalism,
by 
\begin{equation}
D_{0}(\epsilon,\mathbf{p})=\frac{1}{\rho_{m}}\left(\epsilon^{2}+\omega_{T}^{2}+\omega^{2}(p)\right)^{-1}\label{D}
\end{equation}
The minus sign in Eq.(\ref{Vq}) is the most important our observation;
it appears since we deal here with the 2nd order of expansion over
virtual phonons. Factor 4 in Eq.(\ref{Vq}) comes about due to two
variants of pairing in the average $\langle u^{2}(0)u^{2}(x)\rangle$,
and two independent polarizations of TO phonons. We start from the
static interaction potential $V(0,q)$, and will discuss the issue
of frequency dispersion later on.

Integration over $d\epsilon$ is trivial; using Eqs.(\ref{TOspectrum},\ref{g},\ref{Vq})
we come to 
\begin{equation}
V(0,0)=-\hbar\lambda^{2}\frac{(a^{3}\omega_{L}^{2})^{2}}{s^{3}}\int_{\tilde{BZ}}\frac{d^{3}\tilde{p}}{(2\pi)^{3}}\frac{1}{\left[\kappa^{2}+a^{2}\omega^{2}(\tilde{p})/s^{2}\right]^{3/2}}
\label{Vq2}
\end{equation}
where $\tilde{\mathbf{p}}=a\mathbf{p}$ is dimensionless momentum
and $\kappa=a\omega_{T}/s=a/l_{0}$. Dimensionless integral in Eq.(\ref{Vq2})
is equal to $J/2\pi^{2}$, where $J$ is logarithmically large due
to smallness of $\kappa\approx0.12$. We compute $J$ numerically,
and find 
\begin{equation}
V(0,0)=-\hbar\lambda^{2}\frac{(a^{3}\omega_{L}^{2})^{2}}{2\pi^{2}s^{3}}\ln\frac{\eta}{\kappa}\equiv-W\ln\frac{\eta}{\kappa}\label{Vq3}
\end{equation}
where $\eta\approx5.76$. To estimate $T_{c}$ below we will need
to know $V(0,q)$ more accurately, up to the term $\sim q^{2}$. We
find this additional term by expansion over $q$ in the integral (\ref{Vq}).
The resulting integral converges fast at large $p$, so no lattice
cut-off is needed. Finaly we get: 
\begin{equation}
-V(0,q)=W\left[\ln\frac{\eta}{\kappa}-\frac{(ql_{0})^{2}}{12}\right]\label{Vqf}
\end{equation}
Note that dependence on $q$ is relatively weak, as well as dependence
of $V(\epsilon,q)$ on $\epsilon$ (to be discussed below); the reason
is that major (logarithmic) contribution to the integral in Eq.(\ref{Vq})
comes from TO phonons with momenta $p$ in a broad range $k_{F}\leq p\leq\pi/a$.
For the same reason the effects of renormalization of phonon spectrum
due to interaction with electrons are weak at low concentrations $n_{e}\leq1.5\cdot10^{18}cm^{-3}$;
we will discuss these effects later on. 
For completeness, we provide the e-e potential in coordinate space
at $r\geq a$\,($K_1$ is the MacDonald function):
\begin{equation}
-V(r)=\frac{W}{2\pi l_0 r^{2}}K_{1}\left(\frac{2r}{l_{0}}\right)\label{Vr}
\end{equation}

\begin{figure}[H]
\includegraphics[scale=0.4]{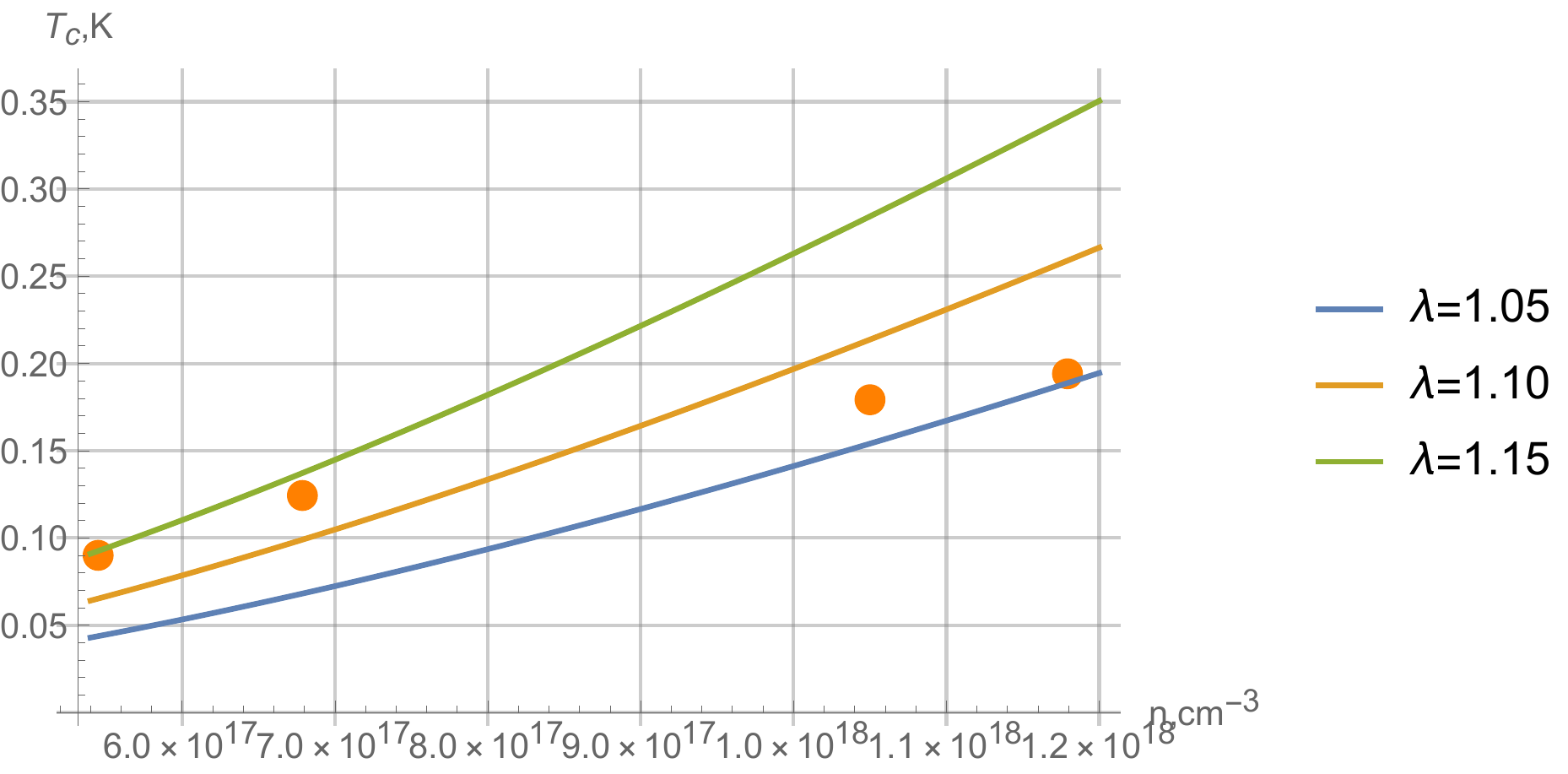}
\caption{\label{fig:Tc(n)} Crtitical temperature as function of conduction electron density
for several values of electron-phonon interaction constant $\lambda$.
Orange points represent experimental data from Ref.\cite{Behnia2014}. }
\end{figure}

\textit{Superconducting transition temperature.} Attractive e-e interaction
defined by Eqs.(\ref{Vqf},\ref{Vr}) decays exponentially at $r>l_{0}/2$
so it can be considered as nearly local in the range of electron densities
$n<n_{c1}=1.5\cdot10^{18}cm^{-3}$. Then we can employ the theory~\cite{Gorkov60}
for superconductivity in a Fermi-gas with local instantenous attraction.
The result~\cite{Gorkov60} for $T_{c}$ is 
\begin{equation}
T_{c}=\zeta\,E_{F}\,\exp\left(-\frac{1}{\nu_{0}V_{0}}\right)\;\qquad\zeta=\frac{e^{C}}{\pi}\left(\frac{2}{e}\right)^{7/3}\approx0.27
\label{Tc1}
\end{equation}
where $\nu_{0}=m_{e}k_{F}/2\pi^{2}\hbar^{2}$ is the DoS per one spin
projection and $V_{0}$ is the $l=0$ harmonics of the pairing potential
(\ref{Vqf}) evaluated at the Fermi-surface. Assuming FS to be spherical
(which is good approximation for STO at low densities), we find (here
$\theta$ is the azimutal angle at the FS, so $q=2k_{F}\sin\frac{\theta}{2}$):
\begin{equation}
V_{0}=\frac{1}{2}\int_{0}^{\pi}|V(0,q)|\sin\theta d\theta=W\left(\ln\frac{\eta}{\kappa}-\frac{1}{6}k_{F}^{2}l_{0}^{2}\right)\label{V0}
\end{equation}
The plot of $T_{c}(n)$ dependence which follows from Eqs.(\ref{Tc1},\ref{V0}),
together with definition of $W$ in Eq.(\ref{Vq3}), is shown in Fig.~\ref{fig:Tc(n)}
together with the data from Ref.~\cite{Behnia2014}.
The choice of electron-phonon coupling constant  $\lambda=1.1$ provides the best correspondence with the data.
Some descrepancy is still present, and it will be discussed below.

\begin{figure}[H]
\includegraphics[scale=0.5]{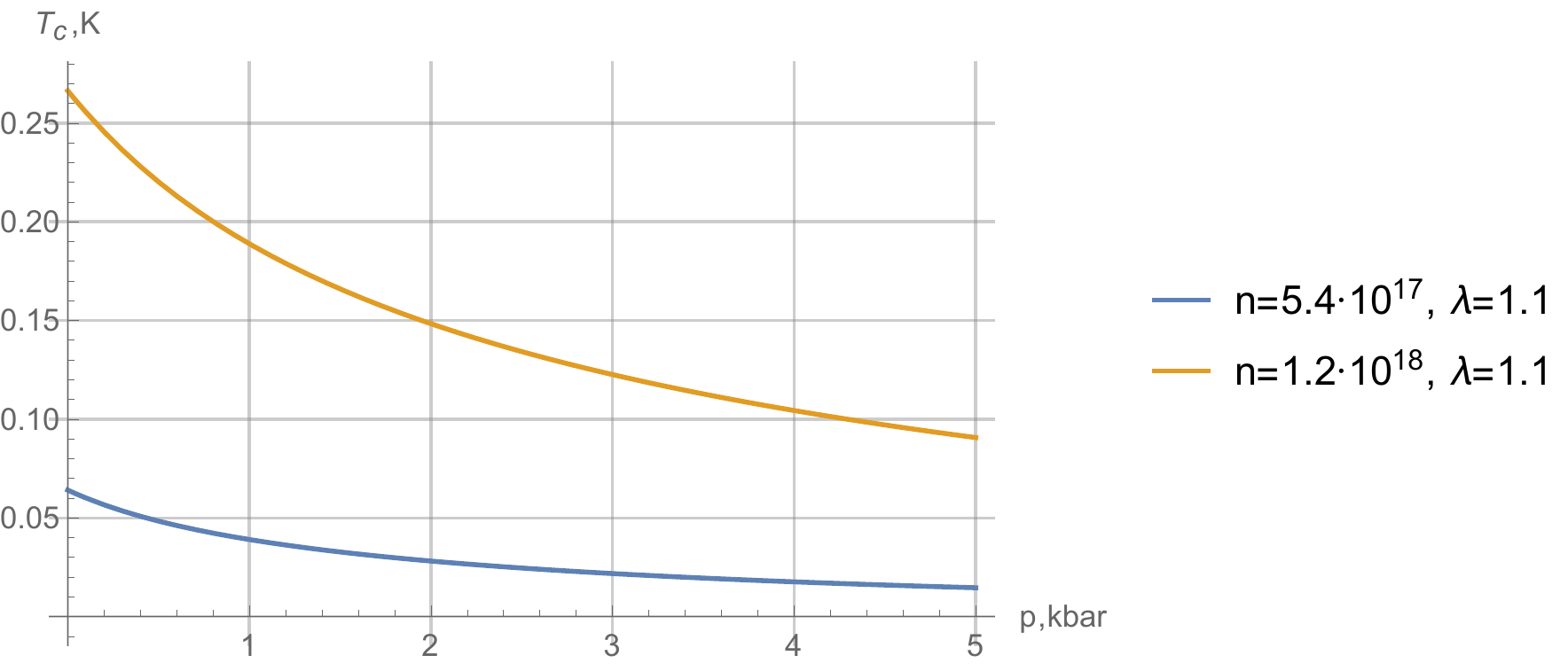}
\caption{\label{fig:Tc(p)}Plot for crtitical temperature-pressure dependence
for two values of electron density indicated in the plot,  obtained
with $\lambda = 1.1$ and  $\omega_{T}\left(P\right)$ dependence extracted from the data of Ref.~\cite{Rowley2020}.}
\end{figure}

\textit{Suppression of $T_{c}$ by hydrostatic pressure.} Relevance
of STO proximity to ferroelectric critical point to the origin of
superconductivity was extensively discussed in Ref.~\cite{Rowley2020}.
In particular, they present data on the effect of hydrostatic pressure
upon $T_{c}$ and upon dielectric constant $\epsilon$, see Fig.2
in Ref.\cite{Rowley2020} which demonstrate that decrease of $\epsilon$
leads to sharp fall of $T_{c}$. According to the standard Lyddane-Sachs-Teller relation,
$\epsilon(P)/\epsilon=(\omega_{T}/\omega_{T}(P))^{2}$ where subsript
$P$ stands for pressure-modified values. Using the data from Fig.2b
of Ref.\cite{Rowley2020}, we calculated, following our theory, 
 $T_{c}(P)$ for our largest electron density $n_{e}=1.2\cdot10^{18}cm^{-3}$.
The result it present in Fig.~\ref{fig:Tc(p)}.
\noindent Accurate comparison of our prediction for $T_c$ suppression with the corresponding data from Ref.~\cite{Rowley2020}
is not possible since they studied the sample with much higher electron density $n= 3.4\cdot 10^{19} cm^{-3}$, the overall
 trend is similar. Our theory predicts a bit smaller suppression effect: factor 2.5 between $P=0$ and $P=$4 KBar,
 while experiment~\cite{Rowley2020} provides supprssion factor close to 4, at the electron density which is 30 times higher.  
One cannot exclude that hydrostatic pressure
may decrease a little the coupling constant $\lambda$, which would lead to additional suppression of $T_c$, not accounted in
our results in Fig.\ref{fig:Tc(p)}.

\textit{Isotopic enhancement of $T_{c}$.} In classical weak-coupling
superconductors with phonon mechanism of e-e attraction, isotopic
substitution of some part of atoms by their heavier isotops leads
usually to weak suppression of $T_{c}$ with increase of the typical
atom mass $M$. The reason is that $T_{c}=1.13\omega_{D}e^{-1/\lambda_{\mathrm{eff}}}$
is proportional to the Debye frequence $\omega_{D}\propto1/\sqrt{M}$,
while effective coupling constant $\lambda_{\mathrm{eff}}$ is independent
on $\omega_{D}$; the latter statement is not entirely evident, but
it follows from Eliashberg theory of phonon-coupled superconductors
~\cite{EliashbergTheory}. Recent experimental data present in Ref.~\cite{Dirk2016}
demonstrate sharp departure from the usual behaviour: substitution
of 35\% of Oxygen atoms $^{16}O$ by their heavy isotop $^{18}O$
resulted in a factor $\approx1.5$ raise of $T_{c}$ for the whole
range of studied electron densities, $4\cdot10^{18}cm^{-3}<n<10^{20}cm^{-3}$.
Another set of data demonstrating the same effect (for higher electron densities)
 can be found in Ref.~\cite{Tomioka}.

\begin{figure}[H]
\includegraphics[scale=0.3]{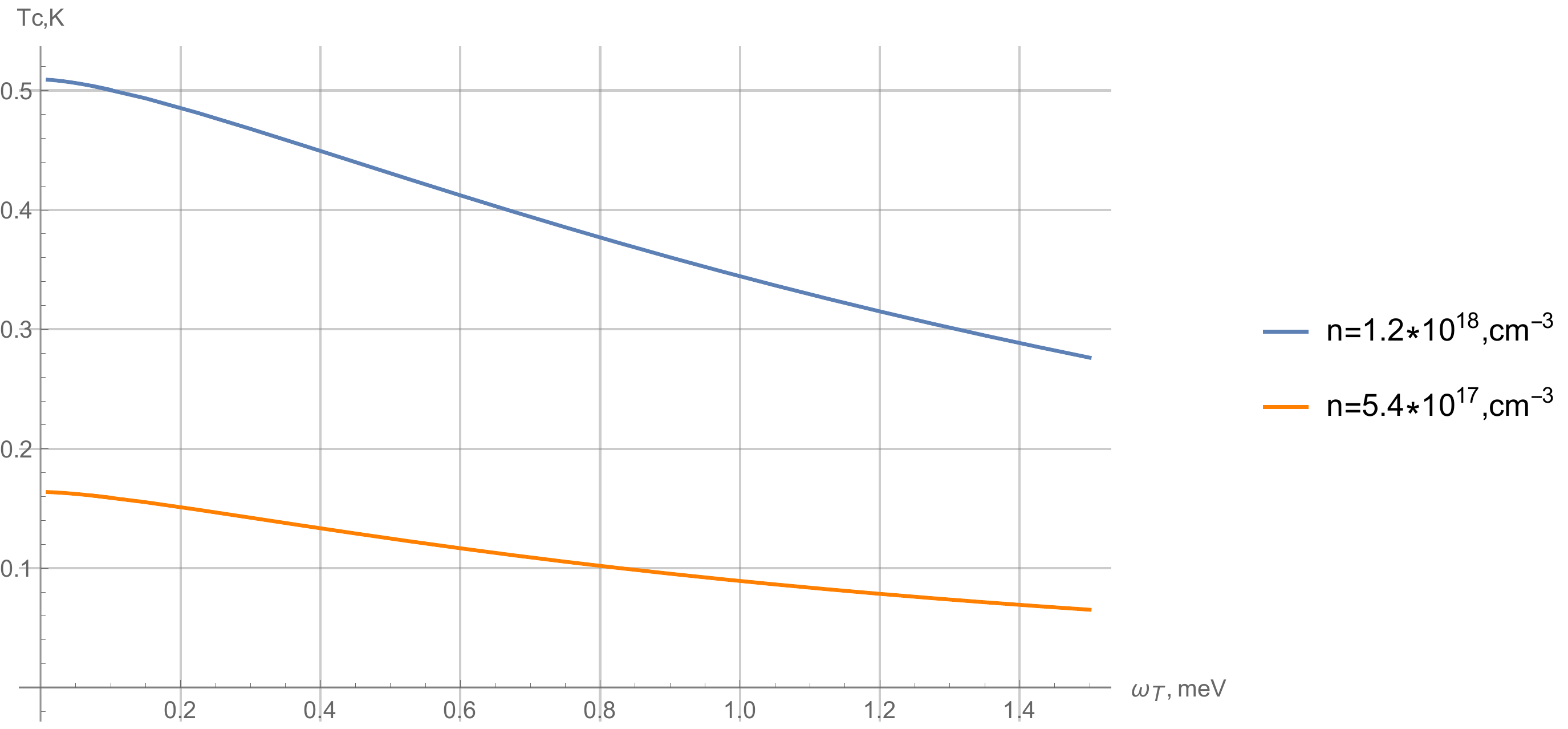}
\caption{\label{fig:Tc(omega)} Crtitical temperature as function of  the TO 
phonon gap $\omega_T$
for two values of the electron density.}
\end{figure}

To explain this giant positive isotop effect, we note that under such
an isotop substitution, insulating STO becomes a ferroelectric~\cite{O18}.
It means that in the isotop-modified material the TO gap $\omega_{T}\to0$.
To take it into account within our theory of superconductivity in
STO, we need just to calculate the ampltiude $V_{0}$ at $\omega_{T}=0$.
Now logarithmic integral in Eq.(\ref{Vq2}) diverges at the lower
limit, while for $V(q)$ we find, instead of Eq.(\ref{Vqf}), the
following result: 
\begin{equation}
\tilde{V}(q)=-W\ln\frac{\eta e}{qa}\label{Vqf2}
\end{equation}
Calculating the $l=0$ scattering amplitude like in Eq.(\ref{V0}),
we find 
\begin{equation}
\tilde{V}_{0}=-W\left(\ln\frac{\eta e^{3/2}}{2k_{F}a}\right)\label{V02}
\end{equation}
We calculated, using Eqs.(\ref{V02}) and (\ref{V0}), the ratio of
transition temperatures for isotop-substituted and native STO at the two values of
carrier concentrations, $n_{\mathrm{low}}= 5.4\cdot10^{17}cm^{-3}$ and $n_{\mathrm{high}} =1.2\cdot 10^{18}cm^{-3}$
 and found
\begin{equation}
\left(\frac{\tilde{T}_{c}}{T_{c}}\right)_{n_{\mathrm{low}}} = 2.5       \qquad  
\left(\frac{\tilde{T}_{c}}{T_{c}}\right)_{n_{\mathrm{high}}} =1.8
\label{ratio}
\end{equation}
The data~\cite{Dirk2016}  provide increase of $T_c$ by factor 1.5 at even higher concentrations, starting from
$n= 4\cdot 10^{18}cm^{-3}$, which seems to be quite consistent with our results in Eq.(\ref{ratio}).
 For a weaker isotopic substitution
$\omega_{T}$ can be only partially suppressed, so the enhancement
of $T_{c}$ will be smaller. We calculated transition temperature $\tilde{T}_{c}$
as function of partially suppressed TO gap $\omega_T$. 
It was done by means of numerical integration starting from $\omega=0$ version of Eq.(\ref{Vq});
 the result is present in Fig.~\ref{fig:Tc(omega)} and can serve
as a prediction for future experiments with isotop-substituted STO.
Note that  relative enhancement of the upper critical field $H_{c2}$ observed in Ref.~\cite{Dirk2016}  due to isotop substitution
is higher than the corresponding  increase of $T_c$ - roughly it is a factor 2 instead of 1.5.
We expect it to be related with additional disorder which accompanies partial isotopic substitution. Indeed,
normal-state resistance of isotop-substituted samples is enhanced, see Ref.~\cite{Tomioka}.

\textit{Conclusions.}
We developed a theory able to predict superconducting transition temperature in lightly doped STO
as function of conduction electron density $n \leq n_{c1} = 1.5\cdot10^{18} cm^{-3}$; the theory has single 
fiting parameter $\lambda$ which determines electron coupling to a pair of TO phonons. Comparision with the data for 
$T_c(n)$ from Ref.~\cite{Behnia2014} selects the optimal value this parameter  $\lambda=1.1$,  
which is close to the value $0.9$ found by totally different method in Ref.~\cite{Nazaryan}.
We note that the estimate for effective strength of e-e attraction $\lambda_{\mathrm{2ph}} \approx 0.28$ found in Ref.~\cite{Dirk2019} for  much 
higher values of $n_e$ does not include large logarithmic factor we discovered, see Eq.(\ref{Vq3}). It is due to integration over large phase
volume of two virtual TO phonons.  Our theory is in good qualitative agreement with experiments on $T_c$ effects due to isotopic
substitution~\cite{Dirk2016,Tomioka} and hydrostatic pressure~\cite{Rowley2020}.

The limitation of low densities used in our theory was used, in the first place, in order to concentrate
on the simplest situation of  a single band filled by conduction electrons, while at higher $n$ second band starts 
to be  filled~\cite{Behnia2014} and more involved calculations are necessary.  
There are few effects which we neglected so far
while they are present, in princple, in the single-band problem as well. 
First of them is the renormalization
of effective phonon gap $\omega_T$ due to the presence of the coupling to electrons. Apparently it is given
by replacement $\omega_T^2 \to \omega_T^2 + 2 g n$ which could lead to considerable effect even in our range of $n$.
However, comparison with the data ~\cite{GapShift} for $n > 10^{19}cm^{-3}$ shows much smaller increase of the gap, 
compatible with $\Delta\omega_T^2 \approx 0.3 g n$, which does not lead to any noticeable effect at $n \leq n_{c1}$.
The difference between the data from Ref.~\cite{GapShift} and naive expectations is probably due to the fact that
TO phonons interact both with conduction electrons and  with ion defects (O deficiency or Nb substitution), 
and these defects partially suppresses the increase of the gap  caused by electrons.

The second relevant effect is due to frequency-dependence of the effective e-e interaction, Eq.(\ref{Vq}).
Indeed, static approximation is definitely fine when $E_F < \hbar\omega_T$.  In fact,  $E_F$ starts to exceed $\hbar\omega_T$
already at $n > 7\cdot 10^{17} cm^{-3}$. However, analysis of  Eq.(\ref{Vq}) shows that 
$V(\epsilon,0)-V(0,0)\approx W\frac{\epsilon^2}{3\omega_T^2}$ while $-V(0,0)=W\ln\frac{\eta}{\kappa}\approx 3.8 W$.
 Thus we expect that retardation effects are relatively minor up to $n_{c1}$, while at higher $n$
the theory~\cite{Gorkov60} should be augmented to take them into account; the same is needed for the accurate analysis of $T_c$ 
enhancement due to isotop substitution leading to $\omega_T$ suppression.


A separate comment refers to the preprint~\cite{Terence} where tendency to superconductivity was observed down to 
extremely low electron densities $3.5\cdot 10^{16}cm^{-3}$, with mid-point $T_c \approx 50-70 mK$ weakly dependent
on $n$ below $10^{18}cm^{-3}$. Simultaneously, the absence of diamagnetic shielding expected for any bulk superconducting
 state was reported~\cite{Terence}.  We expect that the origin of both these observations is related with
large-scale variations of  Oxygen vacancy densities over the sample; then resistivity-defined $T_c$ is controlled
by the regions of larger electron densities, while diamagnetic currents are very weak due to percolative nature of
coupling on large scale.

The absence of Coulomb interaction in doped STO makes it rare representative of a superconductor where universal
effect of $T_c$ suppression by disorder~\cite{Finkelstein1994} is not operating. Moreover, it may demonstrate
the opposite effect of $T_c$ enhancement by strong disorder, predicted earlier in bulk~\cite{Feigelman2010} and 
2D~\cite{Burmistrov2012} materials with suppressed Coulomb interaction. It might be possible to reach the necessary
range of strong disorder, $k_F l \sim 1$, by heavy-dose electron irradiation of STO crystal, along the lines of 
Ref.~\cite{irradiation}.  However, such an irradiation may lead to increase of the gap $\omega_T$ and thus decrease
of effective attraction, so $\omega_T$ dependence on irradiation should be controlled.

\textit{Acknowledgments}
We thank Kamran Behnia,  Cl\'ement Collignon, Benoit Fauqu\'e and Marcin Konczykowski for  useful discussions of experimental aspects of STO. We are grateful to  Alexei Ioselevich, Khachatur Nazaryan and Igor Poboiko for their comments on theory side.  This research was supported
by  the RSCF grant \#  21-12-00104.

\end{document}